\documentclass[runningheads,citeauthoryear]{apinv}
\usepackage{epsfig,cite,graphics}
\usepackage[T2A]{fontenc}
\usepackage[cp1251]{inputenc}
\usepackage{lscape}
\usepackage{longtable}

\begin{document}

\title{Multicolor photometric behavior of the young stellar object V1704 Cygni}
\titlerunning{Multicolor photometric behavior of the YSO V1704 Cygni}
\author{Sunay Ibryamov\inst{1}, Evgeni Semkov\inst{2}, Stoyanka Peneva\inst{2}, U\u{g}ur Karadeniz\inst{3}}
\authorrunning{S. Ibryamov et al.}
\tocauthor{S. Ibryamov et al.} 
\institute{Department of Physics and Astronomy, University of Shumen, 115, Universitetska Str., 9700 Shumen, Bulgaria
	\and Institute of Astronomy and National Astronomical Observatory, Bulgarian Academy of Sciences, 72, Tsarigradsko Shose Blvd., 1784 Sofia, Bulgaria
	\and Department of Physics, Izmir Institute of Technology, Urla, 35430 Izmir, Turkey
	\newline
	\email{sibryamov@shu.bg}}
\papertype{Submitted on; Accepted on}	
\maketitle

\begin{abstract}
Results from $BVRI$ photometric observations of the pre-main sequence star V1704 Cyg collected during the time period from August 2010 to December 2017 are presented.
The star is located in the star-forming HII region IC 5070 and it exhibits photometric variability in all-optical passbands.
After analyzing the obtained data, V1704 Cyg is classified as a classical T Tauri star.
\end{abstract}

\keywords{stars: pre-main sequence, stars: variables: T Tauri, star: individual: (V1704 Cyg)}

\section*{1. Introduction}

One of the most frequent types of pre-main sequence (PMS) stars are the low-mass (M $\leq$ 2M$_{\odot}$) T Tauri stars (TTSs).
The first study of the TTSs as a separate class of variables with the prototype star T Tau was made by Joy (1945).
These stars show irregular photometric variability and emission spectra.

TTSs are divided into two sub-classes: classical T Tauri stars (CTTSs) still actively accreting from their massive circumstellar disks and the weak-line T Tauri stars (WTTSs) without evidence of disk accretion (M\'{e}nard \& Bertout 1999).

According to the classification scheme for the light curves of the CTTSs proposed by Ismailov (2005) can clearly distinguish five principal types based on the light curve shape:
(Type I): constant mean brightness without changes in the amplitude of the rapid brightness variability;
(Type II): constant mean brightness with changes in the amplitude of the rapid variability;
(Type III): varying mean brightness without changes in the amplitude of the rapid variability;
(Type IV): variations of both the mean brightness and the amplitude of the rapid variability; and
(Type V): the variable is often bright, and rare brightness decreases are observed.
The possible physical mechanisms responsible for the different light curve shapes are given in Ismailov (2005).

The long-term multicolor observations are of particular importance to the find and the classification of PMS stars.
The number of the studied PMS stars is not large.
Each individual study in this direction is important and contributes to the overall shaping of the scientific ideas about the star formation process.

V1704 Cygni (also known as LkH$_\alpha$ 155 and [KW97] 50-45) was included in the list of H$_\alpha$ emission-line stars published by Herbig (1958).
The spectrum of the star taken by Mendoza et al. (1990) shows spectral type K3 and strong hydrogen (W$_{\lambda}$(H$_{\alpha}$)$=$98.8 {\AA} and W$_{\lambda}$(H$_{\beta}$)$=$36.7 {\AA}) and forbidden emission lines.
He I ($\lambda$ 5876 {\AA} and $\lambda$ 7065 {\AA}), [O II] ($\lambda$ 3727-9 {\AA}) and [S II] ($\lambda$ 6717 plus $\lambda$ 6731 {\AA}) emission lines are present.
Guieu et al. (2009) included V1704 Cyg in their list of young stellar object candidates.

Section 2 in the present paper gives information about the telescopes and cameras used to collect the photometric observations.
Section 3 describes the derived results and their interpretation.

\section*{2. Observations and Data reduction}

The $BVRI$ photometric observations of V1704 Cyg were performed from August 2010 to December 2017 with the 2-m Ritchey-Chr\'{e}tien-Coud\'{e} (RCC), the 50/70-cm Schmidt and the 60-cm Cassegrain telescopes administered by the Rozhen National Astronomical Observatory in Bulgaria and the 1.3-m Ritchey-Chr\'{e}tien (RC) telescope administered by the Skinakas Observatory\footnote[1]{Skinakas Observatory is a collaborative project of the University of Crete, the Foundation for Research and Technology, Greece, and the Max-Planck-Institut f{\"u}r Extraterrestrische Physik, Germany.} of the University of Crete in Greece. The total number of the nights used for observations is 212.

The observations were performed with four different types of CCD cameras: VersArray 1300B at the 2-m RCC telescope, ANDOR DZ436-BV at the 1.3-m RC telescope, FLI PL16803 at the 50/70-cm Schmidt telescope, and FLI PL09000 at the 60-cm Cassegrain telescope.
The technical parameters and specifications for the cameras used are given in Ibryamov et al. (2015).
All frames were taken through a standard Johnson$-$Cousins $BVRI$ set of filters.
The frames are dark (bias) frame subtracted and flat field corrected.
The photometric data were reduced using \textsc{idl} based \textsc{daophot} subroutine.
As a reference, the $BVRI$ comparison sequence of eleven stars in the field around V2492 Cyg reported in Ibryamov et al. (2018) was used.
All data were analyzed using the same aperture, which was chosen to have a 4$\arcsec$ radius, while the background annulus was taken from 9 to 14$\arcsec$.
The average value of the errors in the reported magnitudes is 0.01$-$0.02 mag for the $I$- and $R$-band data and 0.01$-$0.03 mag for the $V$- and $B$-band data.

\section*{3. Results and Discussion}

V1704 Cyg is located in the field of the Pelican Nebula (IC 5070) at 1$\arcmin$ from the well-studied young star V2492 Cyg.
The results from our long-term $BVRI$ photometric monitoring of V1704 Cyg are summarized in Tab. 1\footnote[2]{The table is available also via CDS Vizier Online Data Catalog.}.
The available photometric data of the star are plotted on Fig. 1.

\begin{figure}[]
  \begin{center}
    \centering{\epsfig{file=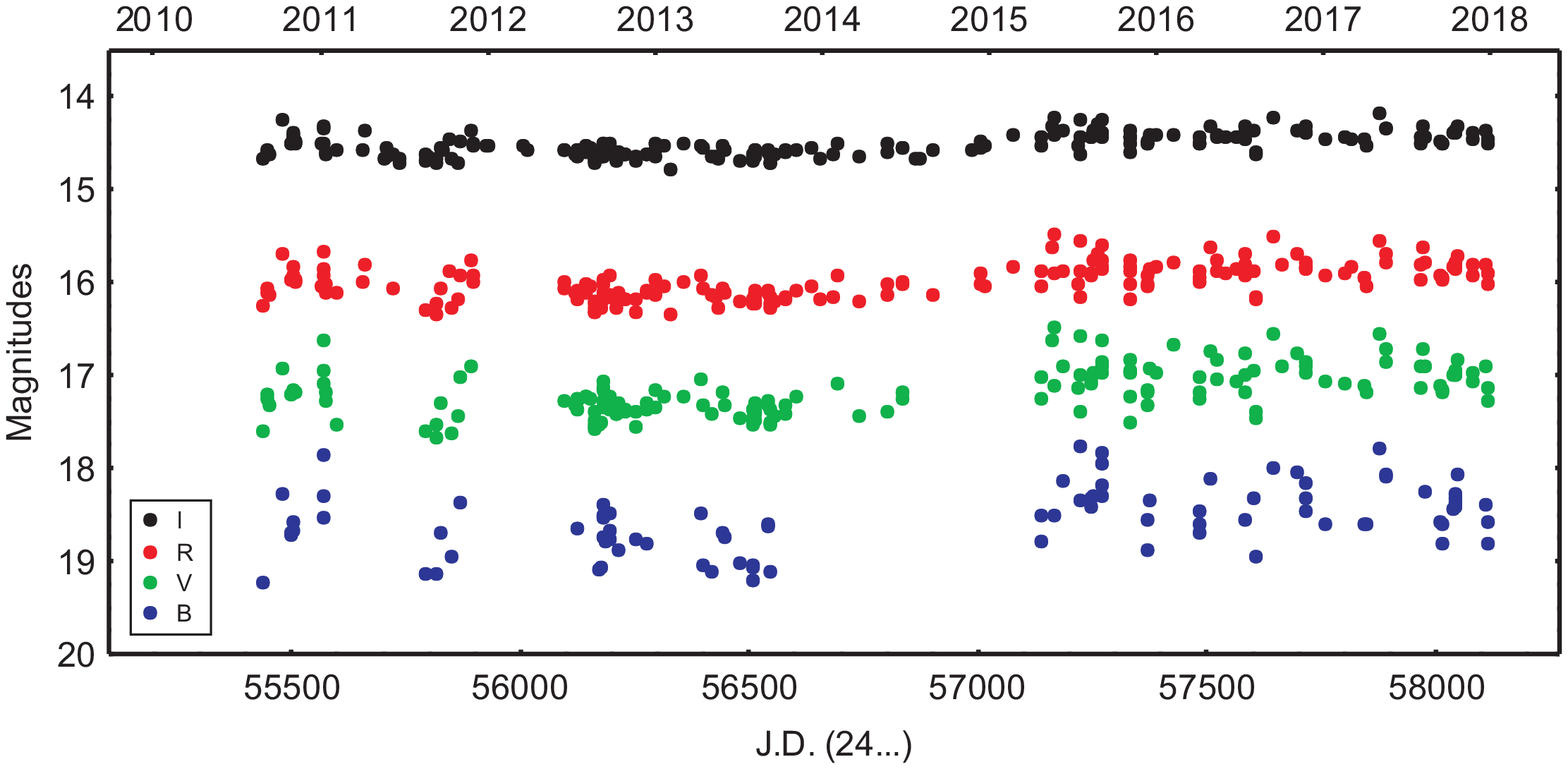, width=\textwidth}}
    \caption[]{$BVRI$ light curves of V1704 Cyg for the time period August 2010$-$December 2017.}
    \label{fig1}
  \end{center}
\end{figure}

{\footnotesize
\begin{longtable}{cccccccc}
\caption{Photometric CCD observations of V1704 Cyg.}\\
\hline\hline
\noalign{\smallskip}  
Date \hspace{1.0cm} &	J.D. (24...) \hspace{1mm}	&	$I$ [mag]\hspace{2mm} & $R$ [mag]\hspace{2mm} & $V$ [mag]\hspace{2mm} & $B$ [mag]\hspace{2mm} & Telescope \hspace{1.0mm} & CCD\\
\noalign{\smallskip}  
\hline
\endfirsthead
\caption{Continued.}\\
\hline\hline
\noalign{\smallskip}  
Date \hspace{1.0cm} &	J.D. (24...) \hspace{1mm}	&	$I$ [mag]\hspace{2mm} & $R$ [mag]\hspace{2mm} & $V$ [mag]\hspace{2mm} & $B$ [mag]\hspace{2mm} & Telescope \hspace{1.0mm} & CCD\\
\noalign{\smallskip}  
\hline
\noalign{\smallskip}  
\endhead
\hline
\label{Tab1}
\endfoot
\noalign{\smallskip}
26.08.2010	&	55435.339	&	14.66	&	16.24	&	17.61	&	19.24	&	1.3-m	&	AND	\\
07.09.2010	&	55447.457	&	14.60	&	16.11	&	17.20	&	-	&	Sch	&	FLI	\\
08.09.2010	&	55448.298	&	14.57	&	16.07	&	17.25	&	-	&	Sch	&	FLI	\\
09.09.2010	&	55449.378	&	14.62	&	16.14	&	17.32	&	-	&	Sch	&	FLI	\\
11.10.2010	&	55481.316	&	14.24	&	15.69	&	16.92	&	18.27	&	1.3-m	&	AND	\\
29.10.2010	&	55499.210	&	14.51	&	15.98	&	17.20	&	18.71	&	2-m	&	VA	\\
30.10.2010	&	55500.184	&	14.50	&	15.98	&	17.20	&	18.69	&	2-m	&	VA	\\
31.10.2010	&	55501.178	&	14.46	&	15.98	&	-	&	-	&	Sch	&	FLI	\\
31.10.2010	&	55501.253	&	14.47	&	15.97	&	17.19	&	18.67	&	2-m	&	VA	\\
01.11.2010	&	55502.178	&	14.48	&	15.97	&	17.15	&	18.57	&	2-m	&	VA	\\
02.11.2010	&	55503.188	&	14.39	&	15.82	&	-	&	-	&	Sch	&	FLI	\\
03.11.2010	&	55504.219	&	14.41	&	15.92	&	-	&	-	&	Sch	&	FLI	\\
04.11.2010	&	55505.175	&	14.44	&	15.98	&	-	&	-	&	Sch	&	FLI	\\
05.11.2010	&	55506.206	&	14.47	&	15.98	&	17.18	&	-	&	Sch	&	FLI	\\
06.11.2010	&	55507.207	&	14.50	&	15.99	&	-	&	-	&	Sch	&	FLI	\\
01.01.2011	&	55563.188	&	14.49	&	16.04	&	-	&	-	&	Sch	&	FLI	\\
06.01.2011	&	55568.169	&	14.31	&	15.67	&	16.62	&	17.85	&	2-m	&	VA	\\
08.01.2011	&	55570.171	&	14.49	&	15.93	&	17.09	&	18.52	&	2-m	&	VA	\\
09.01.2011	&	55571.173	&	14.34	&	15.85	&	16.95	&	18.30	&	2-m	&	VA	\\
11.01.2011	&	55573.215	&	14.62	&	16.02	&	17.18	&	-	&	2-m	&	VA	\\
12.01.2011	&	55574.240	&	14.59	&	16.10	&	17.28	&	-	&	2-m	&	VA	\\
07.02.2011	&	55599.675	&	14.58	&	16.12	&	17.52	&	-	&	Sch	&	FLI	\\
04.04.2011	&	55656.484	&	14.56	&	16.00	&	-	&	-	&	Sch	&	FLI	\\
08.04.2011	&	55659.502	&	14.37	&	15.80	&	-	&	-	&	2-m	&	VA	\\
21.05.2011	&	55703.401	&	14.66	&	-	&	-	&	-	&	Sch	&	FLI	\\
23.05.2011	&	55705.398	&	14.54	&	-	&	-	&	-	&	Sch	&	FLI	\\
25.05.2011	&	55707.379	&	14.63	&	-	&	-	&	-	&	Sch	&	FLI	\\
08.06.2011	&	55721.357	&	14.61	&	16.07	&	-	&	-	&	2-m	&	VA	\\
21.06.2011	&	55734.405	&	14.66	&	-	&	-	&	-	&	Sch	&	FLI	\\
23.06.2011	&	55736.468	&	14.72	&	-	&	-	&	-	&	Sch	&	FLI	\\
16.08.2011	&	55790.302	&	14.62	&	-	&	-	&	-	&	1.3-m	&	AND	\\
17.08.2011	&	55791.306	&	14.68	&	16.29	&	17.61	&	19.14	&	1.3-m	&	AND	\\
23.08.2011	&	55797.310	&	14.64	&	-	&	-	&	-	&	Sch	&	FLI	\\
10.09.2011	&	55815.428	&	14.72	&	16.33	&	17.66	&	19.14	&	1.3-m	&	AND	\\
11.09.2011	&	55816.415	&	14.66	&	16.22	&	17.53	&	-	&	1.3-m	&	AND	\\
19.09.2011	&	55824.258	&	14.55	&	16.06	&	17.29	&	18.69	&	1.3-m	&	AND	\\
23.09.2011	&	55828.268	&	14.57	&	-	&	-	&	-	&	Sch	&	FLI	\\
07.10.2011	&	55842.288	&	14.46	&	15.88	&	-	&	-	&	1.3-m	&	AND	\\
13.10.2011	&	55848.252	&	14.67	&	16.28	&	17.62	&	18.94	&	1.3-m	&	AND	\\
29.10.2011	&	55864.210	&	14.70	&	16.17	&	17.43	&	-	&	2-m	&	VA	\\
31.10.2011	&	55866.256	&	14.47	&	15.93	&	17.02	&	18.36	&	2-m	&	VA	\\
26.11.2011	&	55892.192	&	14.35	&	15.77	&	16.91	&	-	&	2-m	&	VA	\\
29.11.2011	&	55895.188	&	14.49	&	15.92	&	-	&	-	&	Sch	&	FLI	\\
30.11.2011	&	55896.201	&	14.53	&	15.99	&	-	&	-	&	Sch	&	FLI	\\
29.12.2011	&	55925.190	&	14.52	&	-	&	-	&	-	&	Sch	&	FLI	\\
01.01.2012	&	55928.172	&	14.53	&	-	&	-	&	-	&	Sch	&	FLI	\\
17.03.2012	&	56003.552	&	14.52	&	-	&	-	&	-	&	Sch	&	FLI	\\
29.03.2012	&	56015.527	&	14.57	&	-	&	-	&	-	&	2-m	&	VA	\\
15.06.2012	&	56094.409	&	14.58	&	16.07	&	17.27	&	-	&	2-m	&	VA	\\
17.06.2012	&	56096.374	&	14.57	&	16.00	&	-	&	-	&	Sch	&	FLI	\\
11.07.2012	&	56120.350	&	14.62	&	16.10	&	17.32	&	-	&	Sch	&	FLI	\\
12.07.2012	&	56121.313	&	14.64	&	16.17	&	17.36	&	-	&	Sch	&	FLI	\\
13.07.2012	&	56122.375	&	14.57	&	16.13	&	17.25	&	-	&	Sch	&	FLI	\\
14.07.2012	&	56123.347	&	14.56	&	16.08	&	17.28	&	18.64	&	Sch	&	FLI	\\
01.08.2012	&	56141.399	&	14.54	&	16.02	&	17.23	&	-	&	1.3-m	&	AND	\\
02.08.2012	&	56142.285	&	14.53	&	16.02	&	17.22	&	-	&	1.3-m	&	AND	\\
03.08.2012	&	56143.269	&	14.59	&	16.10	&	17.23	&	-	&	1.3-m	&	AND	\\
12.08.2012	&	56151.614	&	14.55	&	16.05	&	17.25	&	-	&	1.3-m	&	AND	\\
19.08.2012	&	56159.326	&	14.64	&	16.20	&	17.39	&	-	&	Sch	&	FLI	\\
20.08.2012	&	56160.303	&	14.69	&	16.32	&	17.58	&	-	&	Sch	&	FLI	\\
21.08.2012	&	56160.540	&	14.66	&	16.26	&	17.56	&	-	&	1.3-m	&	AND	\\
21.08.2012	&	56161.323	&	14.70	&	16.32	&	17.54	&	-	&	Sch	&	FLI	\\
22.08.2012	&	56162.315	&	14.66	&	16.23	&	17.48	&	-	&	Sch	&	FLI	\\
02.09.2012	&	56173.310	&	14.64	&	16.23	&	17.52	&	19.08	&	1.3-m	&	AND	\\
03.09.2012	&	56174.274	&	14.64	&	16.27	&	17.51	&	19.07	&	1.3-m	&	AND	\\
07.09.2012	&	56178.280	&	14.63	&	16.18	&	17.35	&	18.74	&	1.3-m	&	AND	\\
08.09.2012	&	56179.449	&	14.55	&	16.10	&	17.28	&	-	&	1.3-m	&	AND	\\
09.09.2012	&	56180.281	&	14.50	&	15.96	&	17.07	&	18.40	&	1.3-m	&	AND	\\
10.09.2012	&	56181.267	&	14.52	&	16.02	&	17.16	&	18.52	&	1.3-m	&	AND	\\
11.09.2012	&	56182.222	&	14.50	&	16.00	&	17.13	&	18.51	&	1.3-m	&	AND	\\
12.09.2012	&	56183.346	&	14.59	&	16.15	&	17.34	&	18.78	&	1.3-m	&	AND	\\
22.09.2012	&	56193.254	&	14.59	&	16.13	&	17.27	&	18.68	&	1.3-m	&	AND	\\
22.09.2012	&	56193.317	&	14.60	&	16.16	&	17.36	&	18.77	&	Sch	&	FLI	\\
23.09.2012	&	56194.301	&	14.50	&	15.93	&	17.22	&	18.48	&	Sch	&	FLI	\\
07.10.2012	&	56208.229	&	14.69	&	16.27	&	17.42	&	-	&	Sch	&	FLI	\\
08.10.2012	&	56209.232	&	14.65	&	16.21	&	17.42	&	-	&	Sch	&	FLI	\\
09.10.2012	&	56210.214	&	14.65	&	16.20	&	17.42	&	-	&	Sch	&	FLI	\\
11.10.2012	&	56212.241	&	14.59	&	16.11	&	17.29	&	-	&	60-cm	&	FLI	\\
13.10.2012	&	56214.223	&	14.60	&	16.14	&	17.37	&	18.88	&	2-m	&	VA	\\
25.10.2012	&	56226.298	&	14.61	&	16.19	&	17.37	&	-	&	Sch	&	FLI	\\
26.10.2012	&	56227.406	&	14.61	&	16.17	&	17.40	&	-	&	Sch	&	FLI	\\
17.11.2012	&	56249.191	&	14.69	&	16.32	&	17.56	&	-	&	Sch	&	FLI	\\
18.11.2012	&	56250.201	&	14.63	&	16.18	&	17.38	&	18.76	&	Sch	&	FLI	\\
12.12.2012	&	56274.189	&	14.62	&	16.12	&	17.37	&	-	&	2-m	&	VA	\\
14.12.2012	&	56276.191	&	14.62	&	16.08	&	17.29	&	18.81	&	2-m	&	VA	\\
31.12.2012	&	56293.225	&	14.57	&	16.12	&	17.35	&	-	&	Sch	&	FLI	\\
01.01.2013	&	56294.205	&	14.49	&	15.96	&	17.15	&	-	&	60-cm	&	FLI	\\
03.01.2013	&	56296.253	&	14.63	&	16.13	&	-	&	-	&	60-cm	&	FLI	\\
19.01.2013	&	56312.197	&	14.52	&	16.04	&	17.23	&	-	&	2-m	&	VA	\\
05.02.2013	&	56329.182	&	14.78	&	16.33	&	-	&	-	&	Sch	&	FLI	\\
06.03.2013	&	56357.621	&	14.51	&	16.00	&	17.22	&	-	&	60-cm	&	FLI	\\
12.04.2013	&	56394.505	&	14.52	&	15.92	&	17.05	&	18.48	&	Sch	&	FLI	\\
18.03.2013	&	56396.585	&	14.55	&	16.06	&	17.31	&	19.05	&	2-m	&	VA	\\
04.05.2013	&	56417.438	&	14.64	&	16.14	&	17.41	&	19.11	&	2-m	&	VA	\\
15.05.2013	&	56428.423	&	14.64	&	16.15	&	-	&	-	&	60-cm	&	FLI	\\
17.05.2013	&	56430.426	&	14.67	&	16.26	&	-	&	-	&	60-cm	&	FLI	\\
30.05.2013	&	56443.403	&	14.52	&	16.06	&	17.19	&	18.70	&	Sch	&	FLI	\\
31.05.2013	&	56444.375	&	14.58	&	16.12	&	17.31	&	18.73	&	Sch	&	FLI	\\
04.07.2013	&	56478.389	&	14.68	&	16.20	&	17.46	&	19.01	&	2-m	&	VA	\\
01.08.2013	&	56506.376	&	14.68	&	16.21	&	17.48	&	19.05	&	2-m	&	VA	\\
02.08.2013	&	56507.386	&	14.66	&	16.23	&	17.54	&	19.20	&	2-m	&	VA	\\
03.08.2013	&	56508.395	&	14.63	&	16.18	&	17.47	&	19.06	&	2-m	&	VA	\\
04.08.2013	&	56509.310	&	14.61	&	16.18	&	17.37	&	-	&	Sch	&	FLI	\\
05.08.2013	&	56510.365	&	14.62	&	16.19	&	17.49	&	-	&	Sch	&	FLI	\\
05.08.2013	&	56510.392	&	14.60	&	16.14	&	17.40	&	-	&	60-cm	&	FLI	\\
06.08.2013	&	56511.432	&	14.65	&	16.23	&	17.42	&	-	&	60-cm	&	FLI	\\
07.08.2013	&	56512.379	&	14.60	&	16.12	&	17.32	&	-	&	Sch	&	FLI	\\
07.08.2013	&	56512.423	&	14.56	&	16.11	&	17.32	&	-	&	60-cm	&	FLI	\\
08.08.2013	&	56513.401	&	14.62	&	16.20	&	17.33	&	-	&	60-cm	&	FLI	\\
09.08.2013	&	56514.369	&	14.57	&	16.09	&	17.30	&	-	&	60-cm	&	FLI	\\
04.09.2013	&	56540.298	&	14.58	&	16.08	&	17.32	&	18.63	&	Sch	&	FLI	\\
05.09.2013	&	56541.292	&	14.59	&	16.10	&	17.27	&	18.60	&	Sch	&	FLI	\\
07.09.2013	&	56543.426	&	14.71	&	16.23	&	17.52	&	19.12	&	2-m	&	VA	\\
08.09.2013	&	56544.280	&	14.72	&	16.26	&	17.50	&	-	&	2-m	&	VA	\\
11.09.2013	&	56547.408	&	14.64	&	16.16	&	17.36	&	-	&	60-cm	&	FLI	\\
17.09.2013	&	56553.262	&	14.61	&	16.20	&	17.43	&	-	&	1.3-m	&	AND	\\
11.10.2013	&	56577.327	&	14.57	&	16.15	&	17.32	&	-	&	60-cm	&	FLI	\\
12.10.2013	&	56578.352	&	14.59	&	16.18	&	17.42	&	-	&	60-cm	&	FLI	\\
07.11.2013	&	56604.286	&	14.58	&	16.09	&	17.22	&	-	&	60-cm	&	FLI	\\
09.12.2013	&	56636.209	&	14.55	&	16.05	&	-	&	-	&	2-m	&	VA	\\
29.12.2013	&	56656.202	&	14.66	&	16.19	&	-	&	-	&	Sch	&	FLI	\\
23.01.2014	&	56681.208	&	14.61	&	16.16	&	-	&	-	&	Sch	&	FLI	\\
06.02.2014	&	56694.641	&	14.49	&	15.93	&	17.08	&	-	&	2-m	&	VA	\\
22.03.2014	&	56738.568	&	14.63	&	16.21	&	17.43	&	-	&	Sch	&	FLI	\\
21.05.2014	&	56799.450	&	14.51	&	16.01	&	-	&	-	&	Sch	&	FLI	\\
23.05.2014	&	56801.411	&	14.60	&	16.13	&	17.40	&	-	&	2-m	&	VA	\\
23.06.2014	&	56832.439	&	14.54	&	16.00	&	17.24	&	-	&	2-m	&	VA	\\
25.06.2014	&	56834.400	&	14.55	&	16.01	&	17.19	&	-	&	2-m	&	VA	\\
25.07.2014	&	56864.345	&	14.66	&	-	&	-	&	-	&	Sch	&	FLI	\\
03.08.2014	&	56873.383	&	14.67	&	-	&	-	&	-	&	2-m	&	VA	\\
29.08.2014	&	56899.292	&	14.58	&	16.13	&	-	&	-	&	1.3-m	&	AND	\\
26.11.2014	&	56988.172	&	14.58	&	-	&	-	&	-	&	Sch	&	FLI	\\
13.12.2014	&	57005.216	&	14.54	&	16.01	&	-	&	-	&	Sch	&	FLI	\\
14.12.2014	&	57006.246	&	14.48	&	15.91	&	-	&	-	&	Sch	&	FLI	\\
24.12.2014	&	57016.190	&	14.52	&	16.03	&	-	&	-	&	2-m	&	VA	\\
21.02.2015	&	57074.619	&	14.40	&	15.83	&	-	&	-	&	Sch	&	FLI	\\
23.04.2015	&	57136.465	&	14.42	&	15.88	&	17.02	&	18.50	&	Sch	&	FLI	\\
25.04.2015	&	57138.472	&	14.52	&	16.05	&	17.24	&	18.79	&	Sch	&	FLI	\\
18.05.2015	&	57161.391	&	14.32	&	15.62	&	16.63	&	-	&	Sch	&	FLI	\\
21.05.2015	&	57164.440	&	14.23	&	15.48	&	16.47	&	-	&	Sch	&	FLI	\\
24.05.2015	&	57167.386	&	14.41	&	15.90	&	17.11	&	18.50	&	2-m	&	VA	\\
13.06.2015	&	57187.356	&	14.37	&	15.87	&	16.91	&	18.14	&	2-m	&	VA	\\
16.07.2015	&	57220.341	&	14.53	&	16.01	&	17.13	&	-	&	Sch	&	FLI	\\
17.07.2015	&	57221.393	&	14.61	&	16.16	&	17.40	&	-	&	Sch	&	FLI	\\
19.07.2015	&	57223.366	&	14.43	&	15.88	&	16.99	&	18.35	&	2-m	&	VA	\\
20.07.2015	&	57224.394	&	14.25	&	15.54	&	16.57	&	17.77	&	2-m	&	VA	\\
11.08.2015	&	57246.372	&	14.44	&	15.92	&	17.08	&	18.41	&	1.3-m	&	AND	\\
12.08.2015	&	57247.299	&	14.43	&	15.91	&	17.03	&	18.33	&	1.3-m	&	AND	\\
17.08.2015	&	57252.305	&	14.36	&	15.77	&	16.98	&	18.30	&	2-m	&	VA	\\
24.08.2015	&	57259.466	&	14.38	&	15.74	&	-	&	-	&	Sch	&	FLI	\\
25.08.2015	&	57260.364	&	14.29	&	15.68	&	-	&	-	&	Sch	&	FLI	\\
03.09.2015	&	57269.350	&	14.38	&	15.77	&	16.90	&	17.96	&	Sch	&	FLI	\\
04.09.2015	&	57270.433	&	14.38	&	15.85	&	16.86	&	18.18	&	2-m	&	VA	\\
05.09.2015	&	57271.411	&	14.42	&	15.83	&	16.96	&	18.29	&	2-m	&	VA	\\
06.09.2015	&	57272.406	&	14.25	&	15.60	&	16.62	&	17.83	&	2-m	&	VA	\\
03.11.2015	&	57330.215	&	14.35	&	15.75	&	16.82	&	-	&	Sch	&	FLI	\\
04.11.2015	&	57331.201	&	14.44	&	15.87	&	16.96	&	-	&	Sch	&	FLI	\\
05.11.2015	&	57332.204	&	14.59	&	16.17	&	17.51	&	-	&	Sch	&	FLI	\\
06.11.2015	&	57333.202	&	14.51	&	16.02	&	17.23	&	-	&	Sch	&	FLI	\\
07.11.2015	&	57334.199	&	14.42	&	15.84	&	16.94	&	-	&	Sch	&	FLI	\\
12.12.2015	&	57369.186	&	14.48	&	15.93	&	17.16	&	18.56	&	2-m	&	VA	\\
13.12.2015	&	57370.163	&	14.51	&	16.04	&	17.32	&	18.89	&	2-m	&	VA	\\
14.12.2015	&	57371.177	&	14.49	&	16.01	&	17.19	&	-	&	2-m	&	VA	\\
15.12.2015	&	57372.193	&	14.41	&	15.85	&	16.92	&	18.35	&	Sch	&	FLI	\\
02.01.2016	&	57390.173	&	14.41	&	15.83	&	16.97	&	-	&	Sch	&	FLI	\\
07.02.2016	&	57426.188	&	14.40	&	15.78	&	16.67	&	-	&	Sch	&	FLI	\\
04.04.2016	&	57483.498	&	14.43	&	15.87	&	17.02	&	18.46	&	2-m	&	VA	\\
06.04.2016	&	57484.539	&	14.49	&	16.00	&	17.25	&	18.69	&	2-m	&	VA	\\
06.04.2016	&	57485.499	&	14.51	&	15.95	&	17.18	&	18.60	&	Sch	&	FLI	\\
27.04.2016	&	57506.420	&	14.31	&	15.63	&	16.74	&	18.10	&	Sch	&	FLI	\\
13.05.2016	&	57522.406	&	14.40	&	15.75	&	16.84	&	-	&	Sch	&	FLI	\\
14.05.2016	&	57523.408	&	14.44	&	15.87	&	17.04	&	-	&	Sch	&	FLI	\\
31.05.2016	&	57540.384	&	14.43	&	15.90	&	-	&	-	&	2-m	&	VA	\\
25.06.2016	&	57565.442	&	14.44	&	15.85	&	17.06	&	-	&	Sch	&	FLI	\\
11.07.2016	&	57581.363	&	14.38	&	15.80	&	17.00	&	-	&	Sch	&	FLI	\\
12.07.2016	&	57582.391	&	14.46	&	15.93	&	17.18	&	18.56	&	Sch	&	FLI	\\
13.07.2016	&	57583.377	&	14.31	&	15.70	&	16.76	&	-	&	Sch	&	FLI	\\
01.08.2016	&	57602.356	&	14.37	&	15.87	&	16.94	&	18.33	&	2-m	&	VA	\\
04.08.2016	&	57605.360	&	14.60	&	16.15	&	17.38	&	18.95	&	Sch	&	FLI	\\
05.08.2016	&	57606.353	&	14.62	&	16.17	&	17.46	&	-	&	Sch	&	FLI	\\
11.09.2016	&	57643.261	&	14.23	&	15.50	&	16.56	&	18.00	&	Sch	&	FLI	\\
02.10.2016	&	57664.230	&	-	&	15.81	&	16.91	&	-	&	Sch	&	FLI	\\
05.11.2016	&	57698.260	&	14.35	&	15.68	&	16.77	&	18.05	&	Sch	&	FLI	\\
21.11.2016	&	57714.222	&	14.38	&	15.86	&	16.97	&	18.46	&	2-m	&	VA	\\
22.11.2016	&	57715.204	&	14.38	&	15.78	&	16.86	&	18.16	&	2-m	&	VA	\\
23.11.2016	&	57716.215	&	14.31	&	15.83	&	16.89	&	18.31	&	2-m	&	VA	\\
02.01.2017	&	57756.202	&	14.46	&	15.92	&	17.06	&	18.60	&	Sch	&	FLI	\\
17.02.2017	&	57801.603	&	14.42	&	15.89	&	17.09	&	-	&	Sch	&	FLI	\\
05.03.2017	&	57817.549	&	14.45	&	15.83	&	-	&	-	&	Sch	&	FLI	\\
02.04.2017	&	57845.538	&	14.45	&	15.94	&	17.10	&	18.59	&	Sch	&	FLI	\\
03.04.2017	&	57846.567	&	14.53	&	16.05	&	17.18	&	18.61	&	Sch	&	FLI	\\
01.05.2017	&	57875.483	&	14.18	&	15.55	&	16.54	&	17.79	&	2-m	&	VA	\\
18.05.2017  & 57892.420 & 14.33 & 15.68 & 16.72 & 18.08 & Sch & FLI \\
19.05.2017  & 57893.422 & 14.33 & 15.78 & 16.85 & 18.06 & 2-m & VA  \\
01.08.2017	& 57967.382	& 14.49	& 15.98	& 17.13	& -     &	Sch	& FLI \\
02.08.2017	& 57968.322	& 14.42	& 15.81	& 16.89	& -     &	Sch	& FLI \\
03.08.2017	& 57969.336	& 14.32	& 15.63	& 16.72	& -     &	Sch	& FLI \\
12.08.2017	& 57978.495	& 14.43	& 15.79	& 16.90	& 18.26	& Sch	& FLI \\
14.09.2017	& 58011.268	& 14.48	& 15.92	& 17.12	& 18.57	& Sch	& FLI \\
15.09.2017	& 58012.286	& 14.50	& 15.94	& 17.15	& 18.61	& Sch	& FLI \\
16.09.2017	& 58013.269	& 14.51	& 15.97	& 17.18	& 18.81	& Sch	& FLI \\
12.10.2017	& 58039.223	& 14.38	& 15.82	& 17.00	& 18.43	& Sch	& FLI \\
14.10.2017	& 58041.398	& 14.36	& 15.80	& 16.94	& 18.28	& 2-m &	VA  \\
16.10.2017	& 58043.234	& 14.37	& 15.81	& 16.95	& 18.36	& Sch	& FLI \\
16.10.2017	& 58043.272	& 14.38	& 15.85	& 16.99	& 18.33	& 2-m	& VA  \\
17.10.2017	& 58044.295	& 14.39	& 15.79	& 16.99	& 18.42	& Sch	& FLI \\
18.10.2017	& 58045.385	& 14.31	& 15.71	& 16.83	& 18.07	& Sch	& FLI \\
22.11.2017	& 58080.245	& 14.45	& 15.92	& 17.06 & -	    & Sch	& FLI \\
23.11.2017	& 58081.257	& 14.38	& 15.81	& 16.96 &	-	    & Sch	& FLI \\
21.12.2017	& 58109.266	& 14.35	& 15.81	& 16.89	& 18.39	& Sch	& FLI \\
25.12.2017	& 58113.189	& 14.51	& 16.01	& 17.27	& 18.80	& Sch	& FLI \\
26.12.2017	& 58114.223	& 14.46	& 15.90	& 17.13	& 18.57	& Sch	& FLI \\
\hline \hline
\end{longtable}}

The data indicate that V1704 Cyg shows both active states with high amplitudes and quiet states with lower amplitudes, at the same brightness level.
The registered amplitudes of the irregular brightness variations of the star during the whole time of observations are 1.47 mag for the $B$-band, 1.19 mag for the $V$-band, 0.85 mag for the $R$-band and 0.60 mag for the $I$-band.
Such variability is typical of a CTTS and it could be due to the variations in the mass accretion rate and the modulation of the star's brightness in the presence of hot spots on the stellar surface (see Herbst et al. 1994).
Also, the shape of the light curves of V1704 Cyg is typical of the CTTSs of Type II in the classification scheme of Ismailov (2005).

According to Herbst et al. (1994) the hot spots on the CTTSs are small; in most cases they cover a few percent of the stellar surface area.
The temperatures of the hot spots are typically 10 000 K.
Such temperature is significantly higher than that of the stellar photosphere.
Therefore, even hot spots with small sizes cause a significant modulation in the star's brightness.
The hot spots are interpreted as accretion spots or zones, heated by the infall of a column of matter.
Whereas cool spots may last for hundreds or thousands of rotations, hot spots appear to come and go on a much shorter time scale (Herbst et al. 1994).
Probably during the quiet states of V1704 Cyg with lower amplitudes, the hot spots disappear (or change their sizes).
Another possible cause is the decrease of the mass accretion rate.
It is possible both events occur simultaneously.

2MASS $JHK_{s}$ magnitudes of V1704 Cyg were used to construct a color$-$color ($J-H$ versus $H-K_{s}$) diagram to check whether the star has infrared excess, which is a sure indication for the presence of a circumstellar disk.
Fig. 2(left) shows the location of the main sequence (the black line) and the giant stars (the blue line) from Bessell \& Brett (1988), and the location of the CTTSs (the red line) from Meyer et al. (1997).
A correction to the 2MASS photometric system was performed following the procedure in Carpenter (2001).
It can be seen from the figure that V1704 Cyg lies on the line of the CTTSs, i.e. the star has infrared excess indicating the presence of disk around it.

An important result from the photometric study of V1704 Cyg is the variation of the color indices with the star's brightness.
The measured color indices ($V-I$, $V-R$ and $B-V$) versus the stellar $V$-magnitude during the photometric monitoring of the star are plotted on Fig. 2(right).
A clear dependence can be seen from the figure: the star becomes redder as it fades.
Such color variations are typical for the TTSs, whose the variability is produced by the rotational modulation of spot(s) on the stellar surface.

\begin{figure}
\begin{center}
\includegraphics[width=6cm]{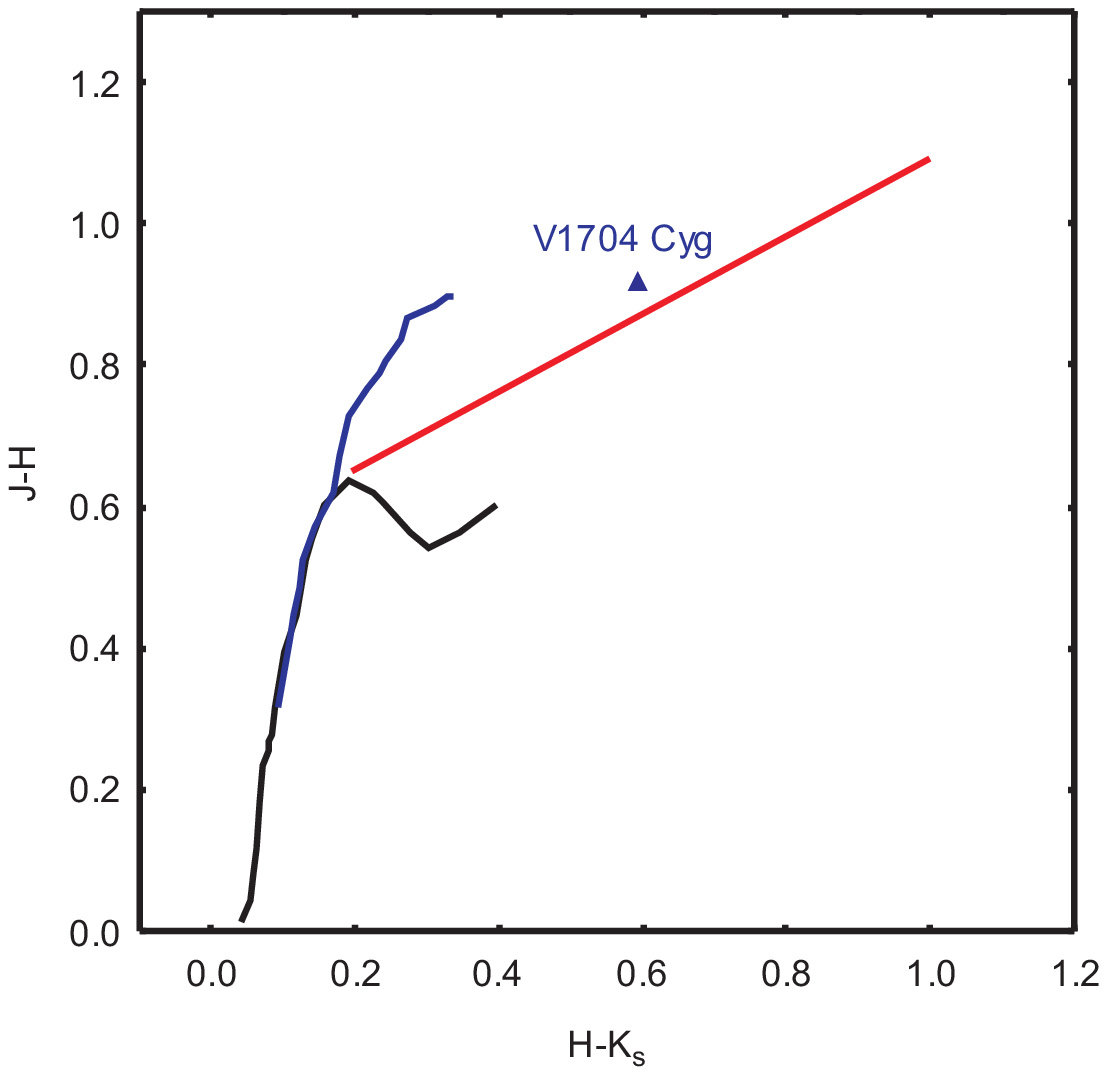}
\includegraphics[width=6cm]{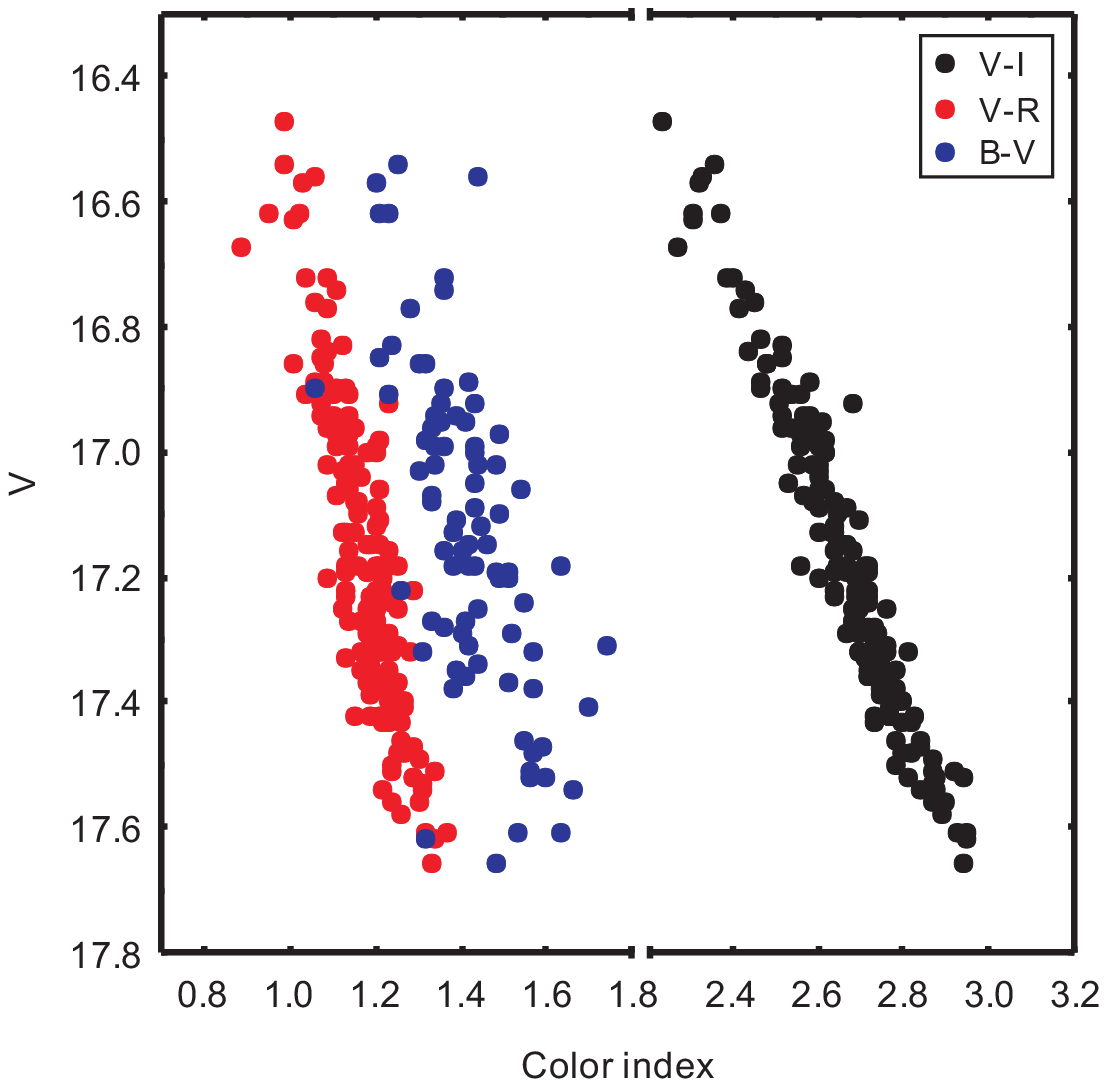}
\caption{(left): Color$-$color diagram for V1704 Cyg detected in the $JHK_{s}$-bands in the 2MASS catalogue; (right): Color indices versus the stellar $V$-magnitude of V1704 Cyg.}\label{Fig2}
\end{center}
\end{figure}

The \textsc{period04} (Lenz \& Breger 2005) software was used to search for periodicity in the light curves of V1704 Cyg.
In Fig. 3 is shown the obtained periodogram using the data in $R$-band.
We found a significant peak in the periodogram corresponding to 1 day period.
This is parasite frequency, an artifact produced by the fact that we observe nightly or at close time-intervals.
The other found peaks correspond to 1/2, 1/3 and 1/4 days overtones of the 1 day alias.
The shapes of the phase-folded light curves according to periods of 1, 1/2, 1/3 and 1/4 days are identical.
Apart from all data obtained during the whole time of observations we used data points from different time periods of observations for time-series analysis.
During these analyzes, we did not find any reliable periodicity in the brightness variations of V1704 Cyg.
The reason is probably the short life of the hot spots or that the variability of the star is not only due to hot spots but also to the variations in the mass accretion rate.

\begin{figure}
\begin{center}
\includegraphics[width=7.0cm]{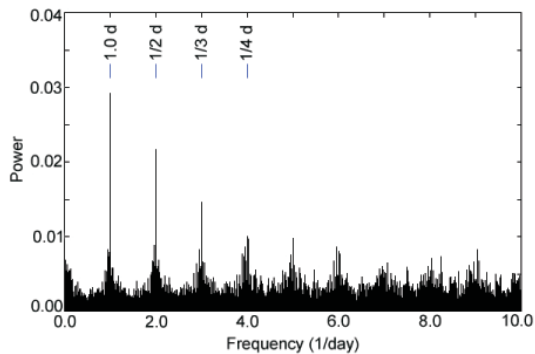}
\caption{Periodogram analysis of timescales inferred from the photometric time-series of V1704 Cyg.}\label{Fig3}
\end{center}
\end{figure}

\section*{4. Conclusion}

The long-term $BVRI$ light curves, the color$-$magnitude and 2MASS two$-$color diagrams of V1704 Cyg were presented and discussed.
We found that the variability of the star is typical of a CTTS.
The shape of the light curves, the location on the two$-$color diagram, the spectrum of the star (reported by Mendoza et al. 1990) and the observed amplitudes in the brightness variations confirmed that conclusion.
Evidence of reliable periodicity in the photometric behavior of V1704 Cyg is not detected.
This study adds a new CTTS to the PMS stars family. 
We are continuing to collect photometric observations of the Pelican Nebula.
In the future, we plan to study several dozen PMS stars located in this star-forming region.

\section*{Acknowledgements}

This research has made use of the NASA's Astrophysics Data System.
This publication makes use of data products from the Two Micron All Sky Survey, which is a joint project of the University of Massachusetts and the Infrared Processing and Analysis Center/California Institute of Technology, funded by the National Aeronautics and Space Administration and the National Science Foundation (Skrutskie et al. 2006).
The authors thank the Director of Skinakas Observatory Prof. I. Papamastorakis and Prof. I. Papadakis for the award of telescope time.
This work was partly supported by the Bulgarian Scientific Research Fund of the Ministry of Education and Science under grants DM 08-2/2016, DN 08-1/2016, DN 08-20/2016 and DN 18-13/2017 and by funds of the project RD-08-112/2018 of the University of Shumen.

\end{document}